\documentstyle{article}
\begin{document}

\begin{center}

{\large {\bf Exact Solutions of Model Hamiltonian Problems with
Effective Interactions}}

\vspace{0.2in}

D. C. Zheng$^{1,2}$, J. P. Vary$^{1,3}$, and B. R. Barrett$^{1,2}$

\end{center}

\thispagestyle{empty}

\vspace{0.2in}
\begin{small}

{\it $^{1}$Institute for Nuclear Theory, University of Washington,
		Seattle, WA 98195}

{\it $^{2}$Department of Physics, University of Arizona, Tucson, AZ 85721}

{\it $^{3}$Department of Physics and Astronomy, Iowa State University,
     Ames, IA 50011}

\end{small}

\vspace{0.5in}

\begin{abstract}
We demonstrate with soluble models how to employ the
effective Hamiltonian approach of Lee and Suzuki to obtain
all the exact eigenvalues of the full Hamiltonian.
We propose a new iteration scheme to obtain the effective
Hamiltonian and demonstrate its convergence properties.
\end{abstract}

\pagebreak

\section{Introduction}

Much effort [1--5] 
has been made in the past few decades
to calculate the shell-model effective interaction from
a realistic nucleon--nucleon (NN) interaction,
which reproduces properties of the deuteron and the NN scattering data.
A completely satisfactory solution to this problem has not
yet been found despite much success \cite{kb,hjorth}.
One of the major difficulties is that the
Rayleigh-Schr\"{o}dinger perturbation series for
the full effective interaction in terms of the nuclear
reaction matrix (G-matrix) \cite{bru} does not
show convergence within the first few orders,
and the degree of effort required to accurately calculate
the higher-order terms grows rapidly \cite{jpv}.
In fact, Schucan and Weidenm\"{u}ller \cite{sw} have shown that in cases
when intruder states are present in the model space,
the perturbation series must diverge.

Another problem which remains largely unsolved is that
shell-model results are found to be dependent on quantities which are,
in principle, arbitrary.
For example, the results are found to be dependent on the unperturbed
single-particle Hamiltonian, $H_{0}$, and on $\omega$,
the starting energy of the
G-matrix. The residual dependence of the results on these quantities
is a measure of the lack of a fully converged effective Hamiltonian.
Of course, one may select a self-consistent $H_{0}$ to
eliminate classes of higher-order diagrams and one
may choose $\omega$ to give reasonable agreement between
theoretical and experimental spectra. However, a fully
converged effective Hamiltonian
would have a strong measure of independence of both $H_{0}$ and
$\omega$.

Some ten years ago, Lee and Suzuki \cite{ls}
proposed an approach to computing the effective
interaction based on a similarity transformation.
They suggested two iteration methods which are able to
assure the independence of the results from the starting energy
and, therefore, to bring a degree of
convergence to the effective interaction.
To employ their approach, however,
it is necessary to evaluate the ``Q-box'', an infinite series
of irreducible, valence-linked diagrams \cite{kuob,qbox}.
Thus, the problem still remains to sum the series that defines the Q-box
to all orders.

The effective Hamiltonian is greatly simplified in a ``no core''
model space \cite{jpvnc}, thereby avoiding some of the convergence
difficulties.
Since there is no inert core, there are no hole lines and the Q-box
reduces, at the two-particle effective-interaction level, to
the bare G-matrix, which has only the ladder diagrams (to all
orders), and can be calculated exactly \cite{bhm,vy}.
The only corrections, in principle, are
many-body forces,
but first one must solve the problem for effective two-body forces
accurately before meaningful statements can be made about these higher-body
forces.  In the no-core approach,
all the particles in a nucleus are active, so for
practical reasons one has to restrict oneself
to light nuclei.

Poppelier and Brussaard \cite{pb} have adopted a no-core model space,
evaluated the G-matrix as the leading contribution to the Q-box,
and have applied the Lee-Suzuki method to light nuclei.
Their work provides some motivation for the present endeavor, since they
obtained encouraging results, even though they
did not achieve $\omega$-independence of the effective Hamiltonian. We
are able to obtain $\omega$-independent effective Hamiltonians
in our applications to soluble models.

We will demonstrate two additional properties of the Lee-Suzuki
approach: independence of the effective Hamiltonian from the choice
of the single-particle Hamiltonian $H_0$ and the ability to obtain
{\it all} the exact eigenvalues of the full problem.
These demonstrations will be given with soluble models.

The organization of this paper is as follows:
We first give a brief review of
the Lee-Suzuki formalism and its application
by Poppelier and Brussaard. We show the independence of the resulting
effective Hamiltonian from $H_0$.
We then suggest a different
iteration equation for the effective interaction
than the one used in Ref.\cite{pb}
and show that when the truncated model space is one-dimensional,
this iteration equation can be reduced to
a simple, intuitively correct expression, stating that the total energy is
equal to the unperturbed energy plus the interaction energy. This
leads to the demonstration that all
the converged effective energies are exact eigenenergies of the
system.
We finally apply our procedure to some simple systems, investigate
the convergence properties of the approach, and show that
{\it all} the exact energies are obtained in the truncated model space.

\section{$H_{\rm eff}$: Method and Properties}
In this section, we give a brief review of the Lee-Suzuki formalism
\cite{ls,qbox,pb} and adapt it to our version of the nuclear no-core
problem, point out the $H_0$ independence and introduce an alternative
iteration scheme.

Let us consider the eigenvalue problem
\begin{equation}
	H \Psi = E \Psi,
\end{equation}
where $H$ is the full Hamiltonian and $\Psi$ is the wave function
of an $A$-particle system. The Hamiltonian $H$ consists of the
relative kinetic energy term
$T_{\rm rel}$ (i.e., the center of mass kinetic energy
has been removed) \cite{trel} and the potential term $V$,
which we take to be a purely two-body operator:
\begin{equation}
	H = T_{\rm rel}+ V = H_{0} + (T_{\rm rel}+ V - H_{0})
	  = H_{0} + H_{I},
\end{equation}
where $H_{0}$ is the arbitrary unperturbed Hamiltonian
and $H_{I}=H-H_{0}$.

We introduce the operator $P$ which projects onto the model space
and the operator $Q$ which projects onto the excluded space, so that,
$1=P+Q$, $PQ$=$QP$=0 and $[H_{0},P]$=$[H_{0},Q]$=0.

We then consider a similarity transformation:
\begin{eqnarray}
{\cal H} &=& e^{-S} H e^{S} \nonumber \\
	 &=& P{\cal H}P + P{\cal H}Q + Q{\cal H}P + Q{\cal H}Q,
\end{eqnarray}
where $S$ has the property that $S=QSP$ (hence, $PSP=QSQ=PSQ=0$,
$S^{2}=S^{3}=\cdots=0,$ $e^{S}=1+S$, $e^{-S}=1-S$).
For $P{\cal H}P$  to be a $P$-space  effective interaction $H_{\rm eff}$,
we require
\begin{equation}
Q{\cal H}P = Q  e^{-S} H e^{S} P = 0.    \label{cond}
\end{equation}
It then follows that if $S$ exists,
any $P$-space eigenvalue is also an eigenvalue of
${\cal H}$ and, therefore, of $H$.
Equation (\ref{cond}), which determines $S$, can be rewritten as
\begin{equation}
QH_{I}P + QHQS - SPHP - SPH_{I}QS = 0.   \label{eq4s}
\end{equation}
After this equation is solved for $S$, the effective Hamiltonian is
given by
\begin{equation}
H_{\rm eff} = P{\cal H} P = PHP + PH_{I}QS,
\end{equation}
and the effective interaction by
\begin{equation}
V_{\rm eff} = H_{\rm eff} - PH_{0}P = PH_{I}P + PH_{I}QS.
\end{equation}

It has been shown by Poppelier and Brussaard \cite{pb} that Eq.(\ref{eq4s}),
after a quantity $\omega S$ is added to both sides,
can be rewritten as
\begin{equation}
S = \frac{1}{\omega - QHQ}QH_{I}P -
    \frac{1}{\omega - QHQ}SP(H_{0}-\omega  + V_{\rm eff})P.   \label{eq4s1}
\end{equation}
In order to solve the above equation iteratively for $S$,
one introduces a quantity $Z$:
\begin{equation}
Z = PH_{0}P - \omega P + V_{\rm eff},     \label{veff}
\end{equation}
and defines the generalized G-matrix:
\begin{equation}
G(\omega) = PH_{I}P + PH_{I}Q \frac{1}{\omega - QHQ} QH_{I}P.  \label{gmatrix}
\end{equation}
This generalized G-matrix (referred to as the ``Q-box'' in Ref.\cite{ls})
is an $A$-body operator since $P$ and $Q$ project onto $A$-particle states.
In applications to nuclei, one expects the Brueckner-G-matrix \cite{bru}
to be a leading approximation to Eq.(\ref{gmatrix}), and this is the
approximation invoked in Ref.\cite{pb}.
In our model, we retain the complete generalized G-matrix throughout.

Next one rewrites Eq.(\ref{eq4s1}) as
\begin{equation}
(1+PH_{I}Q \frac{1}{\omega - QHQ} S)Z  = PH_{0}P - \omega P + G(\omega).
\end{equation}
Poppelier and Brussaard suggest an iteration equation for the
solution of $Z$ ($Z_{1}=0, n\geq 2$):
\begin{equation}
Z_{n} = \frac{1}{1-G_{1}-G_{2}Z_{n-1}-G_{3}Z_{n-2}Z_{n-1}
- \cdots - G_{n-1}Z_{2}Z_{3}\cdots Z_{n-1}}P(H_{0}-\omega + G_{0})P
					\label{iter}
\end{equation}
where $G_0$=$G(\omega)$ and
$G_{k}$ is related to the $k$-th derivative of $G(\omega)$:
\begin{equation}
G_{k}(\omega) = \frac{1}{k!}\frac{d^{k}G(\omega)}{d \omega^{k}}.
\end{equation}
If the iteration procedure converges, the effective interaction can
be obtained from Eq.(\ref{veff}).

One can now see that $H_{\rm eff} = Z + \omega P$ is
independent of $H_0$ since $G(\omega)$ depends only linearly on $H_0$,
leading to the cancellation of the $H_0$ term
in $Z_n$, Eq.(\ref{iter}). This
manifest independence of $H_0$ is one of the more appealing
features of this method.

In any practical application of the iteration procedure (\ref{iter}),
we can only include a finite number of derivatives $G_{k}$.
Assuming $G_{k}=0$ for $k>N$, we can rewrite (\ref{iter}) for
$n>N$ as
\begin{equation}
\begin{small}
Z_{n} = \frac{1}{1-G_{1}-G_{2}Z_{n-1}-G_{3}Z_{n-2}Z_{n-1}
- \cdots - G_{N}Z_{n-N+1} Z_{n-N+2}\cdots Z_{n-1}}P(H_{0}-\omega + G_0)P.
\end{small}					\label{iter1}
\end{equation}
It is easy to see that if the iteration is to converge, the converged value
$Z$ has to satisfy the following equation which can be obtained from
Eq.(\ref{iter1}) by taking the limit $n\rightarrow \infty$ and identifying
$Z_{n=\infty}=Z$:
\begin{equation}
Z = \frac{1}{1-G_{1}-G_{2}Z-G_{3}Z^{2}
- \cdots - G_{N}Z^{N-1}}P[H_{0}-\omega + G(\omega)]P.	\label{eq4z}
\end{equation}
We, therefore, obtain an alternative iteration equation:
\begin{equation}
Z_{n} = \frac{1}{1-G_{1}-G_{2}Z_{n-1}-G_{3}Z_{n-1}^{2}
- \cdots - G_{N}Z_{n-1}^{N-1}}P[H_{0}-\omega + G(\omega)]P.
					\label{iter2}
\end{equation}
We should point out that the above iteration equation can also
be applied to a case for which the $k$-th derivative of the G-matrix
does not vanish (but is sufficiently small so the iteration converges)
for any large $k$.

\section{Test Case: One-Dimensional Model Space}

We now show that for the case in which the model space consists
of only one unperturbed basis state (so all the matrices in the
model space become numbers),
Eq.(\ref{eq4z}) can be
reduced to a particularly simple form which is intuitively appealing.
In fact, we have for $N\rightarrow \infty$, that
\begin{equation}
G_{0} + G_{1}Z + G_{2}Z^{2}
+ \cdots + G_{N}Z^{N} \rightarrow G(\omega +Z).
\end{equation}
So Eq.(\ref{eq4z}) becomes ($E_{0}$ is the eigenenergy of the
unperturbed Hamiltonian $PH_{0}P$)
\begin{equation}
Z = \frac{1}{1+[G(\omega)-G(\omega+Z)]/Z}[E_{0}-\omega+G(\omega)]
\end{equation}
which can be easily simplified to yield
\begin{equation}
\omega + Z = E_{0} + G(\omega + Z).			\label{eq4h}
\end{equation}
This is an equation which a converged value of $Z$ must
satisfy for any starting energy $\omega$. It is evident from
Eq.(\ref{eq4h}) that the effective interaction, given by
$V_{\rm eff}=\omega + Z -E_{0}$, is independent of $\omega$.
In other words, Eq.(\ref{eq4h}) shows that for any chosen $\omega$,
$Z$ adjusts so that $\omega +Z$ is independent of that $\omega$.
Remembering that $H_{\rm eff} = \omega +Z$, we note that Eq.(\ref{eq4h})
is simply stating that the total energy $E$=$H_{\rm eff}$ is the
unperturbed energy $E_{0}$ plus the interaction energy $V_{\rm eff}$,
or,
\begin{equation}
E = E_{0} + G(E),		 \label{eq4e}
\end{equation}
after we identify $V_{\rm eff} = G(E)$.
Note that $E$ again is independent of $H_0$ (or $E_0$)
since $G(E)$ has a term, --$E_0$, cancelling the explicit $E_0$
on the right-hand side of Eq.(\ref{eq4e}).

Let us consider a simple example for which $G(\omega)$
has a quadratic dependence on $\omega$:
\begin{equation}
G(\omega) = (g_{0}-E_0) + g_{1} \omega + \frac{1}{2}g_{2} \omega^{2},
\end{equation}
where $g_{0}$, $g_{1}$, and $g_{2}$ are constants independent of
$\omega$. We have for this choice of $G(\omega)$ that
$$G_{0} = G(\omega), \hspace{0.1in}
  G_{1} = g_{1} + g_{2}\omega, \hspace{0.1in}
  G_{2} = g_{2}/2,$$
and $G_{k}=0$ for $k>2$.
Eq.(\ref{eq4z}) for $Z$ then becomes (note cancellation of $E_0$)
\begin{equation}
Z = \frac{1}{1-G_{1}-G_{2}Z}[E_{0}-\omega + G(\omega)].
\end{equation}
The solutions (denoted by $Z_{+}$ and $Z_{-}$) to this equation are
\begin{equation}
Z_{\pm} = \frac{(1-g_{1})\pm \sqrt{(1-g_{1})^{2}-2g_0 g_{2}}}{g_{2}}
- \omega.
\end{equation}
Hence, the resulting energies, given by
$E=H_{\rm eff} = H_0 + V_{\rm eff} = Z_{\pm} + \omega $,
are clearly independent of the starting energy $\omega$ and of $H_0$.
It is easy to verify that these energies $E$ also satisfy
Eq.(\ref{eq4e}).

We claim, as will be shown explicitly in the next section for a couple of
simple cases,  that with the standard definition Eq.(\ref{gmatrix})
for the generalized G-matrix,
Eq.(\ref{eq4e}) for the energy $E$ in the model space
is equivalent to the secular equation for the eigenenergies of the
system in the entire space:
\begin{equation}
{\rm det} (H- EI) = 0,
\end{equation}
where $H$ is the full Hamiltonian in the matrix form and
$I$ is the unit matrix. This will then be sufficient to conclude that
we may obtain all the exact eigenvalues by these procedures.

\section{Application to Exactly Soluble Models}
In this section, we apply Eq.(\ref{eq4e}), an outcome of
the Lee-Suzuki iteration Eq.(\ref{iter}) in the case of
one-dimensional model space, to two exactly soluble models.
We first consider a simple system with only a pair of
unperturbed basis states, $|1\rangle$ and $|2\rangle$.
Our model space
is the space spanned by the state vector $|1\rangle$
and the excluded space is spanned by $|2\rangle$:
$P=|1\rangle \langle 1|$, $Q=|2\rangle \langle 2|$.
The full Hamiltonian
$H$ is then a $2\times 2$ matrix and the unperturbed Hamiltonian
$H_{0}$ is any $2\times 2$ diagonal matrix:
\begin{equation}
H=\left[ \begin{array}{cc} h_{11} & h_{12}
	\\ h_{21} & h_{22} \end{array}\right],
\hspace{0.3in}
H_0=\left[ \begin{array}{cc} a & 0\\ 0 & b \end{array}\right].
\end{equation}
The corresponding generalized G-matrix, defined by Eq.(\ref{gmatrix}), is then
\begin{equation}
G(\omega) = (h_{11}-a) + h_{12}\frac{1}{\omega - h_{22}}h_{21}. \label{gone}
\end{equation}
So the equation for the effective energy, Eq.(\ref{eq4e}), becomes
\begin{equation}
E = h_{11} + \frac{h_{12} h_{21}}{E - h_{22}}
\hspace{0.2in} {\rm or}\hspace{0.2in}
{\rm det}(H-EI) = 0,
\end{equation}
where $I$ is the $2\times 2$ unit matrix.

Similarly for a system with three unperturbed basis states
$|1\rangle$, $|2\rangle$, $|3\rangle$. We have
$P=|1\rangle \langle 1|$,
$Q=|2\rangle \langle 2|  + |3\rangle \langle 3|$.
The full Hamiltonian and the unperturbed Hamiltonian are written as
\begin{equation}
H=\left[ \begin{array}{ccc}
h_{11} & h_{12} & h_{13} \\ h_{21} & h_{22} & h_{23} \\
h_{31} & h_{32} & h_{33} \end{array}\right],
\hspace{0.3in}
H_0=\left[ \begin{array}{ccc}
a & 0 & 0 \\ 0 & b & 0 \\ 0 & 0 & c \end{array}\right].
\end{equation}
The generalized G-matrix is
\begin{equation}
G(\omega) = (h_{11}-a) + \sum_{i,j=2,3} h_{1i}
	\left(\frac{1}{\omega-QHQ}\right)_{ij} h_{j1}
\end{equation}
with
\begin{eqnarray}
\frac{1}{\omega-QHQ} &=& \left[ \begin{array}{cc}
\omega-h_{22} & -h_{23} \\ -h_{32} & \omega - h_{33}
\end{array} \right]^{-1} \nonumber \\
&=& \frac{1}{(\omega-h_{22})(\omega-h_{33})-h_{23}h_{32}}
\left[ \begin{array}{cc}
\omega-h_{33} & h_{23} \\ h_{32} & \omega - h_{22}
\end{array} \right],
\end{eqnarray}
or
\begin{equation}
G(\omega) = (h_{11}-a) + \frac{
h_{12}(\omega-h_{33})h_{21} + h_{12}h_{23}h_{31}
+ h_{13}h_{32}h_{21} + h_{13}(\omega-h_{22})h_{31}}
{(\omega-h_{22})(\omega-h_{33})-h_{23}h_{32}}.          \label{g3}
\end{equation}
So Eq.(\ref{eq4e}) again becomes
\begin{equation}
{\rm det} (H-EI) = 0,
\end{equation}
and independence of $H_0$ and $\omega$ is again guaranteed.

Thus, we have seen by explicit construction, for a one-dimensional
model space, that solving for the exact effective Hamiltonian gives rise
to the entire spectrum by finding all solutions to Eq.(\ref{eq4e}).
Although we have not proven this property for the general
finite dimensional model space problem,
we nevertheless believe it to be true.

\section{Convergence Properties}
In order to examine the different
convergence properties of the iteration equations
(\ref{iter}) and (\ref{iter2}), we arbitrarily make the following
choices for $H$ and $H_0$:
\begin{equation}
H=\left[ \begin{array}{ccc}
1 & -0.5 & -0.25 \\ -0.5 & 3 & -\sqrt{2} \\
-0.25 & -\sqrt{2} & 4 \end{array}\right],
\hspace{0.3in}
H_0=\left[ \begin{array}{ccc}
1 & 0 & 0 \\ 0 & 2 & 0 \\ 0 & 0 & 3 \end{array}\right].
\end{equation}
The generalized G-matrix, given by Eq.(\ref{g3}), becomes
\begin{equation}
G(\omega) = \frac{\alpha}{\omega-2}+\frac{\beta}{\omega-5}  \label{gapp}
\end{equation}
with $\alpha = (9+4\sqrt{2})/48$ and $\beta = (3-2\sqrt{2})/24$.
It has two simple poles at $\omega$=2 and $\omega$=5.
Note that the poles of $G(\omega)$
are controlled by $H$ and are independent of $H_0$.
The exact eigenvalues $E$ for this problem, which satisfy Eq.(\ref{eq4e}),
or
\begin{equation}
E=1+G(E),
\end{equation}
can be solved diagrammatically,
as shown in Figure 1. In this figure, we plot the function
$\omega$  and the function
$1+G(\omega)$. The solutions $E$ are given by
the intersections of the straight line $\omega$ and the curve
segments for $1+G(\omega)$.

In the same figure, we also show the converged value for
the effective Hamiltonian $H_{\rm eff}=E_0+ V_{\rm eff}$ using the
iteration equation Eq.(\ref{iter})
for a wide range of starting energies $\omega$
(dotted line).
It is seen from the figure that all three solutions to Eq.(\ref{eq4e})
can be achieved through the iteration procedure (\ref{iter})
and that the solution which
is reached depends on the starting energy that is chosen.
When the starting energy $\omega$ is smaller than 2 (the position of the
first pole), the converged  energy is identical to the
lowest eigenenergy.
When $\omega$ is in between the two poles, $2< \omega < 5$,
the first excited state energy is obtained
and when $\omega$ is beyond the second pole, $\omega > 5$,
the second excited state energy is obtained.
Note that the exact eigenvalues do not coincide with the poles of the
generalized G-matrix.

The experience with this simple $3\times 3$ example suggests that,
in more realistic applications, one must sweep $\omega$
through a wide range of values to ensure that one obtains
the ground-state eigenvalue.

What is not evident from the figure is how many iterations
are needed to achieve such a convergence. We show that in Table I.
Our criterion for convergence is that the difference in
the effective interaction between the final two iterations is
smaller than $10^{-7}$, i.e.,
\begin{equation}
|V_{\rm eff}(N)-V_{\rm eff}(N-1)| < 10^{-7},  	\label{conv}
\end{equation}
where $N$ is the number of iterations listed in Table I.
It should be pointed out that when Eq.(\ref{iter}) is used,
the number of derivatives ($G_{k}$) to be summed over
in the denominator grows with the iteration
number $n$ (i.e. $k_{\rm max}=n-1$)
but when the iteration Eq.(\ref{iter2}) is used,
the number of derivatives is fixed. For the current application,
we include $G_{k}$ for $k$=1 to $k_{\rm max}$ with
$k_{\rm max}=200$.

One sees from  Table I that for a very wide range of the starting
energy, it takes no more than 15 iterations to reach the converged value
if Eq.(\ref{iter2}) is used. When Eq.(\ref{iter}) is used,
the number of iterations needed to achieve convergence varies
with the starting energy and is generally larger than the
number of iterations for Eq.(\ref{iter2}).
However, as also indicated in the table,
when Eq.(\ref{iter2}) is used, it is quite often the case that
there is no convergence for the starting energy beyond the
first pole $\omega=2.$

It is also obvious from Table I that with the iteration Eq.(\ref{iter}),
the number of iterations (hence, the number of derivatives
of the Q-box) needed for the convergence specified by Eq.(\ref{conv})
is small
when the starting energy is close to the self-consistent value
(i.e., the converged energy $H_{\rm eff}$). It grows rapidly
as the starting energy is shifted away from the self-consistent value.

\section{Conclusions}
Through analysis and applications with soluble models,
we have explicitly shown that the Lee-Suzuki approach to the
effective interaction leads to a final result for the
energy which is independent of the starting energy $\omega$
and independent of the choice of the single-particle
Hamiltonian $H_0$. In addition, with soluble models,
we show the method has the property that it can converge to
any eigensolution of the full Hamiltonian.
We have demonstrated this for  $2\times 2$ and
$3\times 3$ matrix models; however, we believe it to be true
for the general case of an $N\times N$ matrix. For a $3\times 3$
matrix example, we explicitly indicate when the numerical result
for a one-dimensional model space
converges to the ground-state energy and when it converges to each of the
excited-state energies. The state to which the method
converges depends upon the value of
the starting energy $\omega$ relative to the poles in G. We have also
developed a modified iteration scheme which converges more quickly
than the Lee-Suzuki approach but only in the case of the ground-state
energy. These approaches hold great promise for constructing accurate
and meaningful effective-interaction matrix elements for
shell-model calculations. Large scale calculations with the Brueckner
G-matrix computed from different realistic free N-N potentials
are in progress.

\section{Acknowledgment}
We thank P. Brussaard for helpful discussions.
We thank the Department of Energy for partial support
during our stay at the
Institute for Nuclear Theory at the University of Washington.
Two of us (B.R.B. and D.C.Z.) acknowledge
partial support of this work by the National Science Foundation,
Grant No. PHY-9103011. One of us
(J.P.V.) acknowledges partial support by the U.S.
Department of Energy under Grant No. DE-FG02-87ER-40371, Division
of High Energy and Nuclear Physics.

\vspace{0.2in}

\begin{small}

\end{small}

\pagebreak

{\bf Table I}. The convergence properties
of the iteration procedures (\ref{iter}) and (\ref{iter2})
in a simple case for which the G-matrix is one-dimensional and
of the form $G(\omega)$ given by Eq.(\ref{gapp}).
In the table,
$\omega$ is the starting energy which we vary. $N$
is the smallest number of iterations needed to achieve the convergence
defined by the inequality (\ref{conv}). $H_{\rm eff}=Z+\omega$
is the converged
value for the effective Hamiltonian. For the iteration procedure
(\ref{iter2}) we include the derivatives $G_{k}$ to 200-th order
(i.e., $k=1,2,\cdots, 200$). The symbol
``--" in the table means no convergence within
250 iterations. All three converged values for $H_{\rm eff}$
are identical to the exact eigenenergies of the system.
\begin{center}
\begin{small}
\begin{tabular}{rrrr}\hline\hline
$\omega$ \hspace{0.05in}
& $N_{\rm Eq.(\ref{iter})}$ & $N_{\rm Eq.(\ref{iter2})}$
		& $H_{\rm eff}$ \hspace{0.1in} \\ \hline
-20.0  &177 &12 &  0.75338 \\
-10.0  &100 &12 &  0.75337 \\
 -5.0  & 59 &11 &  0.75337 \\
  0.0  & 15 & 8 &  0.75337 \\
  0.75 &  4 & 4 &  0.75337 \\
  1.2  & 19 & 9 &  0.75337 \\
  2.25 &  5 & 5 &  2.24479 \\
  2.5  & 11 &54 &  2.24479 \\
  3.0  & 18 &-- &  2.24479 \\
  4.0  & 43 &-- &  2.24479 \\
  5.1  &  8 &-- &  5.00183 \\
  6.0  & 19 &-- &  5.00183 \\
 10.0  & 51 &-- &  5.00183 \\
 20.0  &127 &-- &  5.00183 \\ \hline\hline
\end{tabular}
\end{small}
\end{center}

\vspace{0.5in}

\section*{Figure Caption}
{\bf Figure 1}. The converged effective Hamiltonian $H_{\rm eff}=Z+\omega$,
with $G(\omega)$ given by Eq.(\ref{gapp}), as a function of the
starting energy $\omega$ (dotted line) using Eq.(\ref{iter}).
The exact solutions for
$H_{\rm eff}$ are given by the intersections of the solid curves
[$\omega$ and $E_0+G(\omega)$], see Eq.(\ref{eq4e}).

\end{document}